\documentstyle[emulateapj]{article}
\input epsf

\begin{document}


\newcommand{\dT}{{\Delta }}              
\newcommand{\lmean} {{\overline{\ell}}}  
\newcommand{\band}{{\rm \tiny ^{BAND}}}


\title{FIRST RESULTS FROM VIPER: DETECTION OF SMALL-SCALE ANISOTROPY AT 40 GHZ}

\author{%
J.~B. PETERSON${}^{1,2}$, G.~S. GRIFFIN${}^1$, M.~G. NEWCOMB${}^1$, 
D.~L. ALVAREZ${}^1$, C.~M. CANTALUPO${}^1$, D.~ MORGAN${}^1$, 
K.~W. MILLER{}$^3$, K. GANGA${}^4$, D.~ PERNIC${}^5$, M.~ THOMA${}^5$  }

\footnotetext[1]{Department of Physics,
	Carnegie Mellon University, 
	Pittsburgh, PA 15213 }

\footnotetext[2]{email:jbp@cmu.edu}

\footnotetext[3]{JILA and Deptartment of Physics, University of Colorado, 
Boulder, CO 80303}

\footnotetext[4]{IPAC, Caltech, Pasadena CA 91125 and
PCC, Coll\`ege de France, F-75231
Paris Cedex 5, France}

\footnotetext[5]{Yerkes Observatory, Williams Bay, WI 53191}


\begin{abstract}

Results of a search for small-scale anisotropy in the cosmic microwave
background (CMB) are presented. Observations were made at the 
South Pole using the Viper telescope, with a $.26^\circ$
(FWHM) beam and a passband centered at 40 GHz. Anisotropy
band-power measurements in bands centered at
$\ell=  108, 173, 237, 263, 422$ and $589$ are reported.
Statistically significant anisotropy is detected in all bands.
\end{abstract}


\keywords{cosmology: cosmic microwave background, observation}


\section{ Introduction }

Most theories of the early universe predict the presence of a
peak in the cosmic microwave background (CMB) anisotropy power spectrum
(\cite{white94}). 
In cold dark matter 
models the position of this peak, the first acoustic peak,  is 
$\ell \sim 220 \Omega_{total}^{1\over2}$ where
$\Omega_{total} = \Omega_{matter} + \Omega_\lambda$
is determined by the total matter/energy density in the universe
(\cite{kam94}). 
In many inflation models 
 $\Omega_{total}$ is forced to one; these 
inflation models make the specific prediction that the
power spectrum will peak at $\ell \sim 220$.

CMB anisotropy data have recently been analyzed in 
comparison to theoretical models (\cite{line98}, \cite{ratra99},
\cite{huweb99}, \cite{scottweb99}) and
there is evidence for a peak in the power spectrum. That 
conclusion has in the past been based on combined analysis of many different
experiments at various angular scales, since most previous individual
experiments did not cover the range of $\ell$
best suited to search for the expected peak.
The Viper telescope, when used at 40 GHz, has a $0.26^\circ$ 
beam (FWHM) which sweeps $3.6^\circ$ across the sky. Thus Viper
has sensitivity from $\ell \sim 100$ to $\ell \sim 600$,
spanning the range of interest for tests of inflation and other
cosmological models.



\section{ Instrument } 

The optics of the Viper telescope consist of four mirrors arranged
in an off-axis configuration.
The 2.15m primary mirror, together with the
secondary mirror, form an aplanatic gregorian.
Radiation from a distant object, converging to a focus
after reflection from the  primary and secondary,
is reflected again by a flat, electrically driven
chopping mirror, and then directed into the photometer feed horns
using a fast hyperbolic condensing mirror. 
The chopping mirror is placed at an exit pupil of the gregorian,
i.e. at an image of the primary formed by the secondary.
This means that tilting the chopping mirror is nearly equivalent,
optically, to tilting the primary.

Because the optical design has a clear aperture,
the secondary and condensing mirrors
can be built oversized without the
blockage that would occur in an on-axis telescope.  We have done this
to improve optical efficiency and reduce
pickup of earth emission.
The primary mirror has incoherent extension 
panels that increase the effective diameter to 3 m, and 
the entire telescope is housed in a 10 meter diameter
conical reflecting baffle to further reduce pickup of earth emission.
Details of the instrument design are available at the Viper web 
page\footnote[6]{http://cmbr.phys.cmu.edu/viper}.

For the observations reported here, the photometer used on Viper was 
a two pixel receiver based on HEMT (high electron mobility transistor) 
amplifiers cooled to 10-20 K temperature. The amplifiers are 
coupled to the telescope through corrugated feed horns chosen to 
provide a half-power illumination pattern on the primary of about 
1 m diameter. This photometer,
called Corona, measures the total power
from 38 to 44 GHz, in two sub-bands.

The instrument is calibrated using ambient temperature and liquid-nitrogen-immersed
calibrators, temporarily inserted so they fill the illumination
pattern of the feed horns. The efficiency of the optics,
$90\pm5$\%, is measured by tracing the beam through the 
telescope using ambient-temperature absorbers, and is checked
by measuring the brightness temperature of the Moon.
The total calibration uncertainty is 8 \%.


\section{ Observations }

The chopping mirror oscillates at a frequency of $2.35$ Hz, causing the beam 
to sweep back and forth across the sky in the co-elevation direction at nearly
constant velocity. Using observations of Venus (figure \ref{beam}), 
we find that the chopper throw is $3.60 \pm .01^\circ$, and the beam is 
$.26 \pm .01^\circ$ wide (FWHM) with a Gaussian shape and
no noticeable eccentricity. This is consistent
with similar measurements made using Centaurus A 
and a remote Gunn oscillator.

\centerline{\null}
\vskip1.65truein
\includegraphics{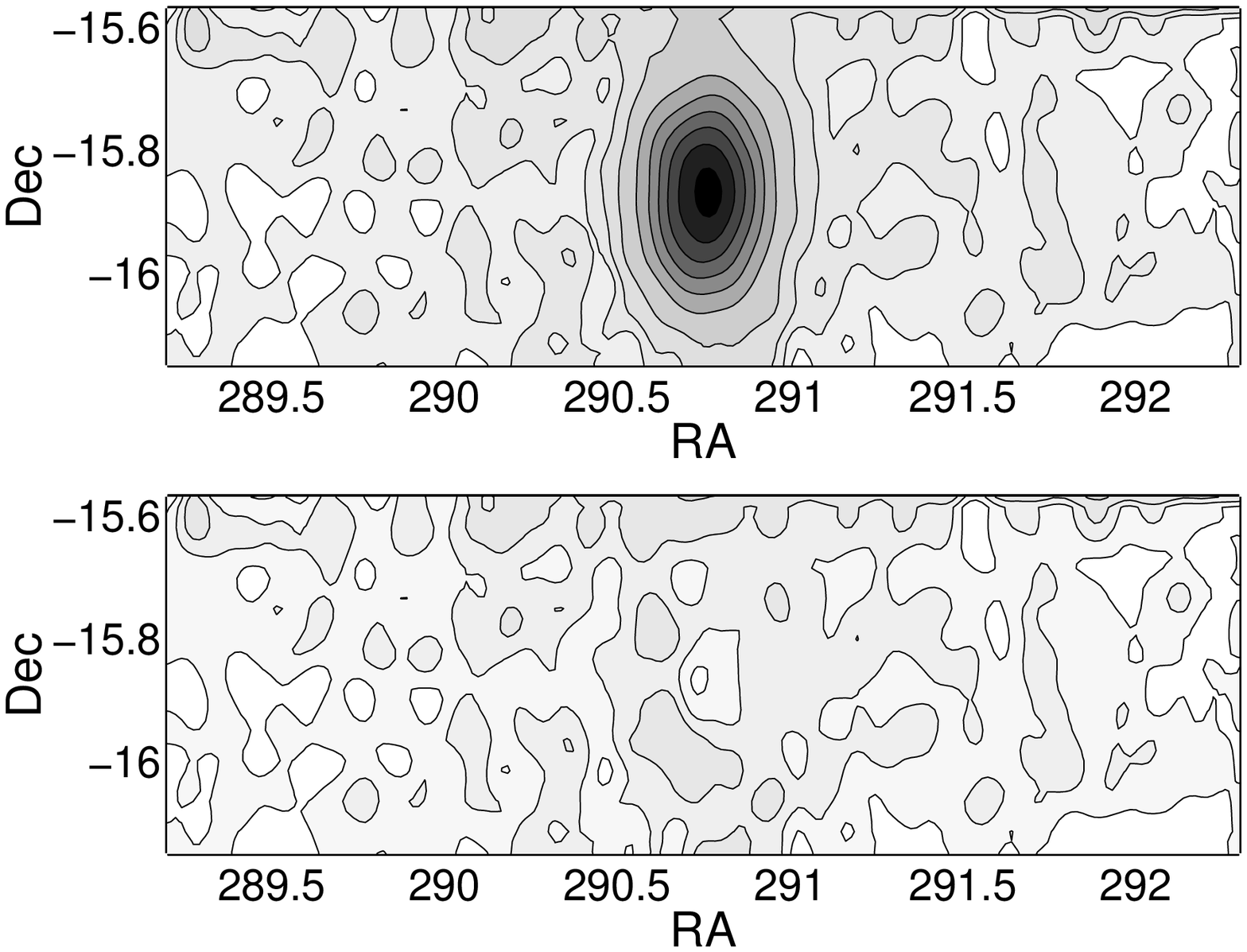}
\figcaption{
Antenna Beam Pattern.   The top image is the beam pattern of the
Viper/Corona system measured using Venus as a source.  The image is created
by  sweeping the beam in co-elevation using the chopping mirror
while stepping through many elevation angles.  We model our
beam pattern as a 0.26$^\circ$ FWHM Gaussian. The bottom panel
shows the fit residual after the model beam pattern has been subtracted from
the data in the top panel.  Contours are at 25, 50, 100, 200...700 mK.
\label{beam}}
\vspace*{0.3cm}

For the observations discussed here,
the telescope slews at a constant declination, $-52.03^\circ$
(epoch 2000.0), 
between 11 overlapping fields spaced 
$.77^\circ$ apart. It dwells at each field for 13.7 s, 
and spends 5.0 seconds slewing to the next field. For fields
$i=1...11$, the movement pattern is: i=1,3,5,7,9,11,10,8,6,4,2.
In this manner,
a single {\it scan} containing $151$ s of data
is completed in $206$ s. 
A total of 135 hours of data, recorded in June 1998, are included in this analysis.

Interspersed with
these observations, the Carina nebula was swept
for a few minutes every 2 hours  
as a pointing  check. Over the period of this data set, the
measurements of the nebula center varied by $< 1$ arcminute (rms). 
We take this to be our relative pointing precision.

To determine the absolute pointing accuracy of the instrument, we scanned 6 bright
($\ge$ 25 mK) objects at declinations 
$-63.0\le\delta\le-47.5$ 
in the galactic plane. 
We use a simple 5-parameter pointing model\footnote[7]{
These parameters are: the offsets of the azimuth and elevation 
encoders, the distance of the two feed horns from the optical center, 
and the dewar angle
in the focal plane. } 
to align these objects with their radio coordinates, determined using 
the Parks-MIT-NRAO (PMN) 4.8 GHz survey (\cite{pmn93}, \cite{pmn94}). We find a 
pointing residual of 2.4 arcminutes rms. Using this pointing
model, we scan 30 Doradus 
($5h38m43s -69^\circ06'03''$) 
and find it 3.8 arcminutes from its PMN coordinates. 
This is the largest discrepancy between the radio 
positions and the positions we find at 40 GHz, so we believe our 
absolute pointing to be accurate to within
4 arcminutes in the region $-69.1\le\delta\le-47.5$.


\section{Synchronous Offset Subtraction}
As the chopping mirror moves, sweeping the beam across the sky, the photometer uses
different areas of the telescope mirrors. 
The illumination of the primary 
mirror hardly changes as the chopping mirror moves, but the
illumination pattern on the secondary changes substantially. The emissivity of the 
secondary might vary across its surface
if, for example, emissive snow had accumulated in an uneven pattern. In addition, 
scattering by snow grains on the optics might vary as the chopper sweeps.  
These effects produce what are called
synchronous offsets, which appear at the detector output to be 
variations of sky brightness
but are in fact instrumental in origin.  If
these offsets  come from the telescope 
they will produce the same apparent sky structure regardless of where the 
telescope is pointed
so they can be identified and removed.  The removal process is described in
the next section.


\section{ Data Reduction }

If any sample deviates by more than 5-sigma from the average, the sweep containing
that sample is deleted. This serves to remove electrical interference 
that appears in a few places in the data. 
If more than 2 sweeps are deleted from any scan, the entire scan
is deleted. 

We then determine the synchronous offset.
For each scan, we co-add over all the sweeps in each field to produce
a single waveform. We then co-add 
all 11 fields to produce an average chopper-synchronous offset waveform for
the scan. This offset waveform, typically less than 4 mK rms in size, 
is subtracted from
the waveform for each field.

Finally we look for rapid changes in the synchronous offset.  
The offset might change if snow is falling on the telescope mirrors, or if clouds
are moving through the scan. We co-add the first and
last 5 fields in each scan. When these two waveforms are subtracted, the
standard deviation of the residual indicates the rate of change of the
offset over the period of a scan. If this exceeds 2.5 mK rms the scan is deleted. This removes a total of 42 hours of data.

\centerline{\null}
\vskip1.95truein
\includegraphics{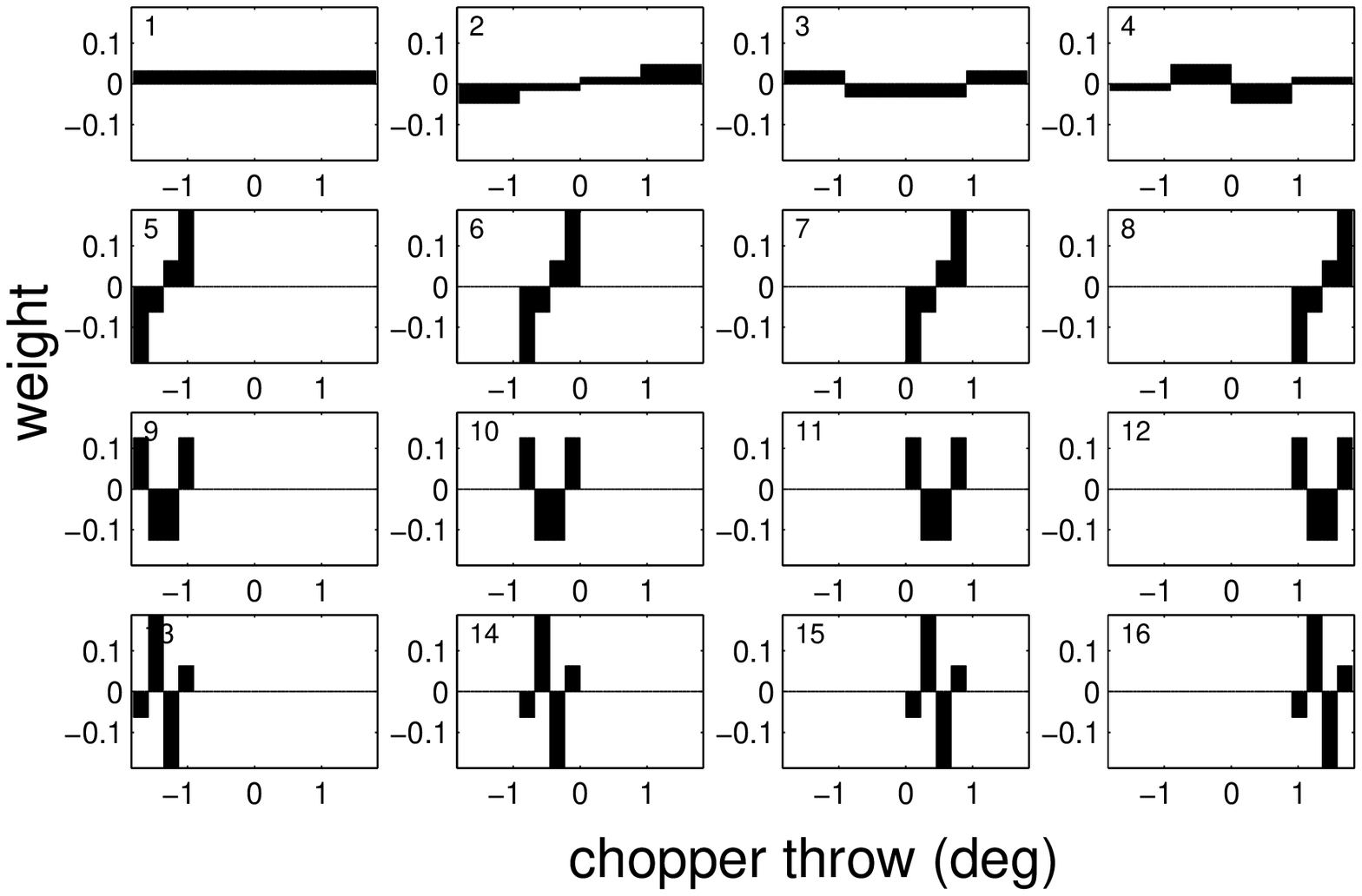}
\figcaption{
Modulation Weighting Functions.   The sixteen modulation weighting functions
used in this analysis are shown versus position within the 3.60 $^\circ$ sweep.
Some patterns differ only by translation.  There are seven unique patterns 
allowing measurement of the anisotropy power spectrum across six ranges
of spatial frequency $\ell$.  The number in the top left of the
upper panels is the index $j$. 
\label{mods}}
\vspace*{0.3cm}


\section{ Modulations }

By multiplying the waveform for each field by the weighting functions
in figure \ref{mods}
we synthesize 16 orthogonal beam patterns. The weighting functions 
$w_{jk}$
are derived using the orthogonality and
normalization conditions:
\begin{eqnarray}
   {\sum_{k=1}^N w_{j_1k} \cdot w_{j_2k} } \, = & 
	\, 0 & \,\,  j_1 \ne j_2,  \,  1 \le j \le N  \\
   {\sum_{k=1}^N | w_{jk} |} \, = & 
	\, 2 
\end{eqnarray}
Here $j$ is an index identifying the function,
and $k$ indexes  $N$ positions across the sky. 
Because fluctuations of atmospheric emission (sky noise) typically have much larger
spatial scales than the region of sky we sweep, the dominant contributions of
the atmospheric emission to our measured signals appear as constant or
gradient waveforms
across the sweep.

To generate the $w_{jk}$, we start with a constant $w_{1k}$ and a gradient $w_{2k}$. Each
successive
modulation $w_{jk}$ is a described by a polynomial of order j-1.
This
constraint, along with equations (1) and (2), generates a unique set of functions that are similar to
the Legendre Polynomials,
except that they 
are defined over a discrete basis. For $j>2$ these weighting functions offer
excellent sky noise rejection because they are orthogonal to
the constant and gradient waveforms.

In this analysis we have divided the sweeps into 4 segments,
and each segment is weighted with $w_{1k}...w_{4k}$. 
We then divide each segment into 4 smaller bins, 
which are similarly weighted.

\centerline{\null}
\vskip3.8truein
\includegraphics{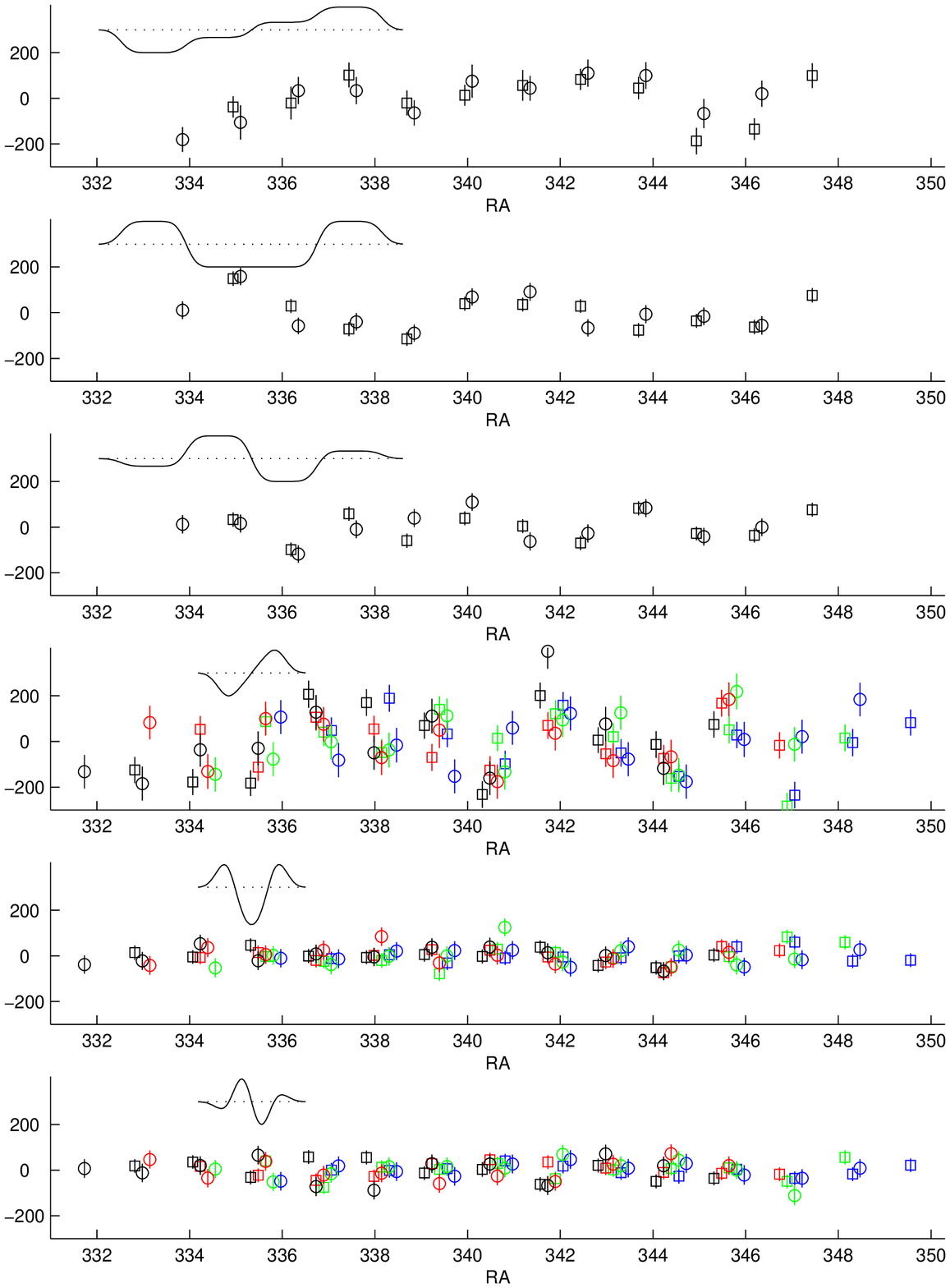}
\figcaption{Temperature Structure on the Sky.
Sky temperature structure is shown for 
six modulations.  The modulation functions, smoothed with the antenna pattern,
are shown in the upper left corner of each plot.
Circles and squares indicate
measurements taken
with the left and right feed horns, respectively. For the last 3
modulations, the chopper throw is divided into 4 measurements each,
denoted by a different colors. Referring to figure 3,
modulations 5, 9 and 13 are black, modulations 6, 10 and 14 are red, 
modulations 7, 11 and 15
are green, and modulations 8, 12 and 16 are blue.
\label{data}}
\vspace*{0.3cm}

Some of the resulting 16 patterns are translated equivalents of others.
There are seven unique patterns. The constant weighting
provides no information on sky structure and is discarded.
We had not expected to be able to detect sky structure with
the gradient weighting $w_{2k}$, but we find that the
South Pole winter atmosphere is so stable that statistically
significant structure is in fact detected.  We continue the analysis
with the six remaining unique weighting functions.
The beam patterns for the six weightings 
are shown in figure \ref{data} and the corresponding window functions are
shown in figure \ref{clplot}.


\section{Band Power Estimates}

The goal of our observations is to measure $C_\ell$, the power spectrum of
CMB anisotropy (see \cite{hu97},  \cite{huweb99}).  
For each modulation we determine the band-power (\cite{bond98}),
using  the matrix method (\cite{net97}).
That is, we generate two matrices, a theory matrix and a noise matrix.

The theory matrix accounts for correlations of modulated sky temperature structure
that would be expected in a noiseless observation of the sky.  
This matrix consists of
correlations calculated between
pairs of modulated observations on separated fields. 
The theory matrix is calculated without reference to the data
from the sky, using lagged window functions. It accounts for the fact that our
fields overlap and that pairs of temperature values for nearby locations
are expected to be correlated.  

In contrast to the theory matrix, the noise matrix is calculated from the set of
measured temperatures of the sky
(\cite{line94}).
There are a number of possible sources of correlated noise in these observations.
For example, on days with patchy cloud cover we see variations of
apparent sky brightness that are actually coming from the cloud pattern 
passing over the telescope (sky noise).  This noise is likely to be correlated
since individual clouds move from field to field. 
These correlations must be accounted for if the uncertainty of the observation is to be
correctly estimated.  The noise matrix accomplishes this since its elements 
are the cross-correlations of
data sets recorded for each field.

To determine the range of band-power values that reasonably fit our data
we carry out a likelihood test for each modulation, establishing a  most likely 
$C_\ell$ value.
We also estimate an uncertainty in $C_\ell$ by determining a confidence interval
containing 67 \% of the integrated likelihood. 
These values are plotted in figure \ref{clplot}. 
In estimating $C_\ell$ we used the method of \cite{church98}
to compensate for the effect of offset subtraction. The error bars in figure \ref{clplot} do not include the calibration uncertainty.

\section{Foregrounds}

Because these observations were made over a small range of observing 
frequency, we make use of
observations made with other telescopes to constrain 
possible astronomical foregrounds.

The observations reported here lie within the 
region previously studied with Python (\cite{coble99}), 
a region which was carefully selected
to have a very low level of foreground emission.  
Using Python the region has been mapped
with $\sim  1^\circ$ angular resolution at 40 GHz and 90 GHz.  These maps show strongly
correlated sky structure. The 
sky structure detected with Python has a frequency spectrum 
consistent with the CMB  and not consistent with
the spectrum of any single known foreground source.
Using data from the Infra-Red Astronomical Satellite (IRAS) (\cite{iras88})
and using PMN data, the Python team has estimated that of the
$\sim$80 $\mu$K  RMS sky structure
detected less than $\sim$ 1 $\mu$K rms is due to
dust, synchrotron or free-free emission. 
Galactic structure has a spatial power spectrum that falls rapidly with $\ell$,
and the Viper observations reported here cover a higher range of $\ell$ than the
Python results.
We therefore accept the Python foreground analysis as indicating that extended
(Galactic) foregrounds make no significant contribution to the
sky structure measured on these fields with Viper.

\centerline{\null}
\vskip2.2truein
\includegraphics{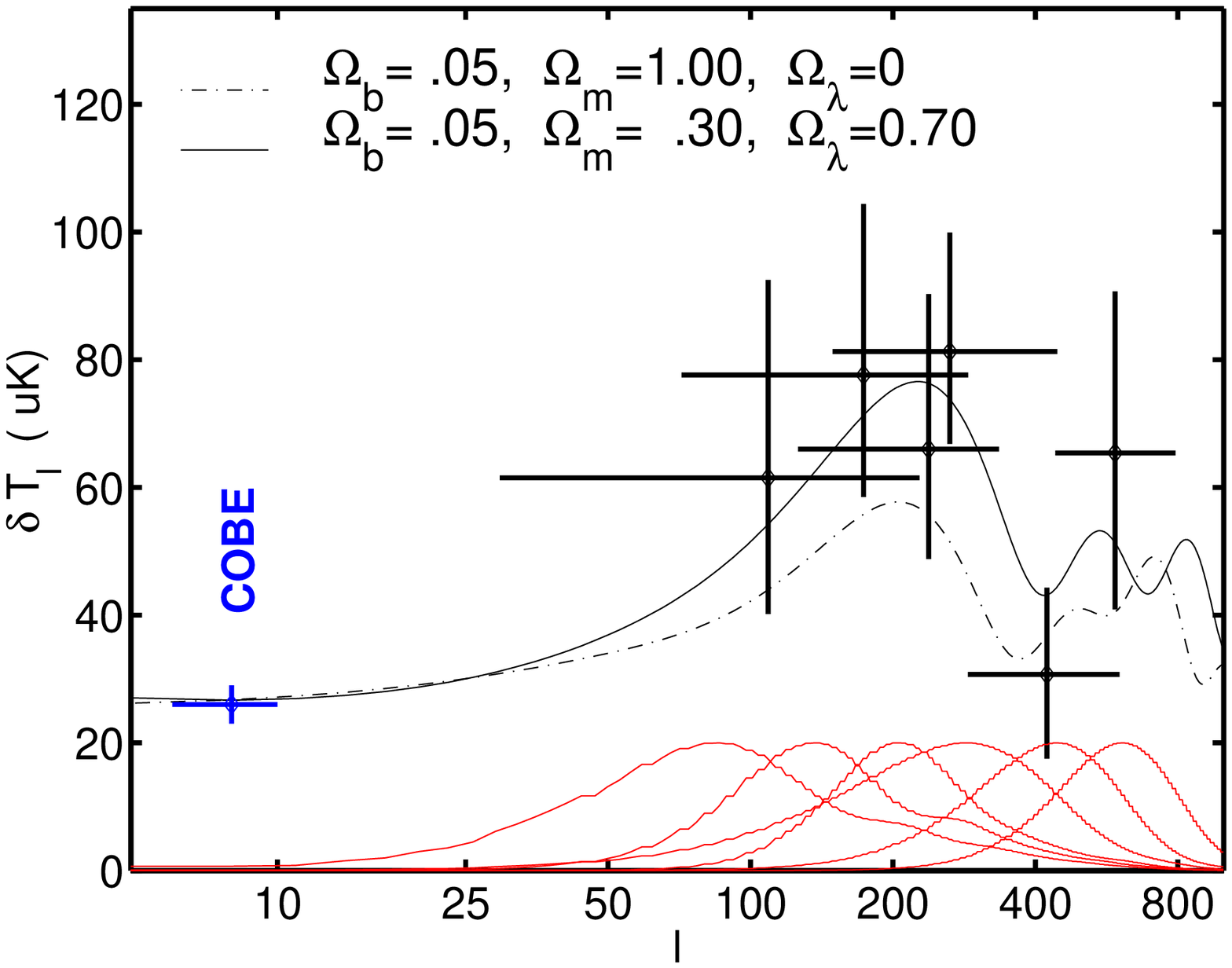}
\figcaption{The Anisotropy Power Spectrum.
The band-power for six modulations is shown as a function of
the spatial frequency $\ell$, along with two
theoretical curves calculated using cmbfast (\cite{cmbfast}).
Also shown is an average COBE band-power data point. At the bottom of the
plot are window functions (scaled to the same height) for each modulation.
\label{clplot}}
\vspace*{0.3cm}

Viper has a beam about 1/16 the area of the Python beam 
so foregrounds due to extra-galactic point sources 
are a more significant concern.  
To estimate the contribution due
to unresolved sources,  
we examined a total of 160 PMN 4.8 GHz sources that fall close to the region we observed.
The brightest source in the vicinity of our observations is PMNJ2256-5158.  This source can be expected 
to contribute substantially to the sky structure we detect because
it has flux 700 $\mu$Jy at 4.8 GHz,
our sweep passes dierctly over it, and its spectrum
has been measured to fall only slowly between 2.8 and 4.8 GHz ($S_\nu  \propto
\nu^\beta ; \beta(2.8..4.8)=-0.7$).
We compared our data for the region near this source to that expected
from 5158 and found a statistically significant match.  
If we treat our measured sky brightnesses in this region as
due entirely to 5158 (i.e. ignoring any CMB structure), 
we can estimate a 
spectral index for this source by comparing our data to the
4.8 GHz PMN flux.  We get $\beta(4.8..40)= -0.9 \pm 0.2$.
Because this falls in line with the PMN spectral index, we elect to
make a correction for this
source, by extrapolating from 4.8 GHz, using spectral index $-0.9$, 
and then removing the
contribution to $C_\ell$ due to 5158.  The correction is small:
the greatest effect is in the high $\ell$ bands and amounts to about 2 $\mu$K.
All other PMN sources in the region are at least five times weaker
and would require spectral index $> 1.0$  to
contribute significantly.
We make no correction for any other PMN source.  
The data presented in figure \ref{data} include the emission of
5158 (i.e. we have applied no correction to this data),
but the $C_\ell$ spectrum in figure \ref{clplot} has
been corrected.

Data taken with
IRAS at 100 $\mu$m (3000 GHz) show eight sources near the
region observed.  Extrapolating from 3000 GHz 
to 40 GHz, using a conservative emissivity law $\epsilon \propto \nu^1$, 
we find that
none of the sources can contribute as much as 1 $\mu$K.  We make no correction for 
IRAS sources.

We can not rule out the possibility that an
undetected population of extra-galactic objects contributes to 
fine scale sky structure at 40 GHz.
Our study of possible extra-galactic foregrounds
assumes that
sources detected at lower frequencies have spectra that fall with
frequency when compared to the CMB, as 
radio-bright galaxies do.
We also assume that the objects detected at 100 $\mu$m with IRAS
have spectra that rise with frequency over this frequency range.
This is indeed the pattern seen in infrared-bright galaxies.
There remains the possibility, however, that a class of
previously undetected objects exists with spectra that
closely mimic the CMB.  As an example
consider a dusty galaxy at redshift $z=10$. At that epoch the
CMB temperature was 30K so dust in such a galaxy might have been heated to just
a few Kelvin above the CMB temperature.  That dust emission,
redshifted to millimeter wavelengths, would have a frequency
spectrum peaking just above the CMB peak and
would be exceedingly difficult to distinguish from CMB structure.
While there is currently no evidence that any such objects did exist at redshift 10,
it will take fine-beam millimeter wavelength
sky survey data to rule out the contribution of such sources.


\section{Discussion}

We interpret the sky structure we have detected as CMB anisotropy.
The increase above COBE anisotropy levels, expected in flat inflation
models to appear near 
$\ell \sim  220$, is evident in the data, as is a 
lower anisotropy level near the 
expected position of the first null at
$\ell \sim 400$.  Models with a significant
cosmological constant appear to fit the data better than those
with $\Lambda= 0$.


\acknowledgements

This work was supported by the National Science Foundation under
cooperative agreement OPP-8920223 with the Center for Astrophysical Research in 
Antarctica and was also supported with
internal CMU funds. We wish to thank Ted Griffith and Laszlo Varga for 
constructing
and installing the Viper telescope. Mark Dragovan and Brian Crone
did the initial optical design, and Hien Nguyen modified this
design. We thank the following for contributions to Viper assembly and 
testing: Pamela Brann, Nicole Cook, Alex van Gaalen, Michael Vincent,
Alex Japour,  
Mike Masterman and 
Mark Williams. We also thank the staff of the Amundsen-Scott research station at the 
South Pole.



%
%



\end{document}